\documentclass[conference]{IEEEtran}

\usepackage{booktabs} 

\usepackage{graphicx}

	\usepackage{amsmath}
	\usepackage{amsfonts}
    \usepackage{longtable}
	\usepackage{fancyhdr}
	\usepackage{mdframed}
\mdfdefinestyle{style1}{leftmargin=1cm,rightmargin=0.5cm,skipbelow=0.5cm,skipabove=0.2cm,
   innertopmargin=2pt}
	\usepackage{balance}
    \usepackage{fancybox}
    \usepackage{booktabs}
	\usepackage{enumerate}
    \usepackage{framed}
	\usepackage{listings}
    \usepackage{subcaption}
    \usepackage{xcolor}
    \usepackage{hyperref}

\ifCLASSOPTIONcompsoc
  \usepackage[nocompress]{cite}
\else
  \usepackage{cite}
\fi

\begin{document}
\title{Towards a Model of Testers' Cognitive Processes: Software Testing as a Problem Solving Approach}


\author{Eduard Enoiu$^{1}$, Gerald Tukseferi$^{1}$, Robert Feldt$^{2}$
\\$^{1}$Software Testing Laboratory, M\"alardalen University, V\"aster\aa s, Sweden%
\\$^{2}$Department of Computer Science and Engineering, Chalmers University of Technology, Gothenburg, Sweden%
}

\IEEEtitleabstractindextext{%
\begin{abstract}
Software testing is a complex, intellectual activity based (at least) on analysis, reasoning, decision making, abstraction and collaboration performed in a highly demanding environment. Naturally, it uses and allocates multiple cognitive resources in software testers. However, while a cognitive psychology perspective is increasingly used in the general software engineering literature, it has yet to find its place in software testing. 
To the best of our knowledge, no theory of software testers' cognitive processes exists. 
Here, we take the first step towards such a theory by presenting a cognitive model of software testing based on how problem solving is conceptualized in cognitive psychology.
Our approach is to instantiate a general problem solving process for the specific problem of creating test cases.
We then propose an experiment for testing our cognitive test design model.
The experiment makes use of verbal protocol analysis to understand the mechanisms by which human testers choose, design, implement and evaluate test cases. 
An initial evaluation was then performed with five software engineering master students as subjects. The results support a problem solving-based model of test design for capturing testers' cognitive processes.
\end{abstract}
}
%
%



\maketitle
\IEEEdisplaynontitleabstractindextext
\IEEEpeerreviewmaketitle

\section{Introduction}
For many years, researchers have tried to create new techniques for effectively testing software. This research has answered numerous questions, but still many of them remain unanswered. For example, not that much research has focused on discovering how humans perform testing \cite{lenberg2015behavioral}. The purpose herein is to develop a model of the cognitive processes involved in software testing. 

A traditional approach to software testing is to manually create test cases or scripts that guide test execution and how the software's results are checked for correctness. When these test cases are not a priori created and used during test execution, the approach is called exploratory testing \cite{whittaker2009exploratory}. In both approaches, humans are using their cognitive abilities to create or explore different scenarios for finding bugs and check if the software meets its requirements. A significant portion of the software testing effort involves the design of test cases, test plans, and strategies based on a variety of test goals. Ammann and Offutt \cite{ammann2016introduction} classified the process of test design in two general approaches: criteria-based test design satisfying certain engineering goals and human-based test design based on domain and human knowledge of testing. However, in practice, this is an artificial distinction, since these are complementary and in many cases, human testers are using both to fully test software. Itkonen et al. \cite{itkonen2009testers} observed the testing sessions of eleven software professionals performing system level functional testing. They identified several practices for test session and execution strategies including exploratory testing, systematic comparison of different software versions and input partitioning. It is obvious that in a domain like software testing, test goals and strategies for creating test cases seem to be less well defined as problems in other areas like physics and math. Research in software testing has identified several variables that influence the quality and performance of software testing \cite{itkonen2007defect,itkonen2012role}. Among these are knowledge and strategies for testing, as well as external factors.

In response to this need to better understand the test design process, this work explores the cognitive processes used by testers engaged in software testing. In this paper, we presume software testing as a problem solving process and map the steps and sequence by which testers perform test activities. As a first step towards investigating if there is value in our approach, we present an experimental design and the results of testing it in a pilot study. We hypothesize that a cognitive model of test design can be represented as a cyclical problem solving process. Empirically evaluating this model requires studying the process used by testers and programmers to test software programs. We are proposing an experiment during which subjects are asked to test a program to the best of their abilities based on a program specification while thinking aloud. The resulting verbal protocols are transcribed, coded and content analysis is then used to validate the hypothesized process steps. We performed an initial pilot study using five students to verify that the participants utilize a pattern of cognitive process steps as well as knowledge acquisition strategies to derive test cases. If our model would be further empirically validated and improved, we argue it can help enhance the test design process as well to better align testing tools with the cognitive needs of testers. Finally, we outline different methods that can be used to research and refine the proposed cognitive model of software testing.

\section{Background}
The software testing process can be divided into four steps \cite{ammann2016introduction}: test design, test automation, test execution, and test evaluation. These activities occur within an organizational environment and one or more persons are assigned to perform them. Many factors influence these activities such as organizational structure, training, experience, testing knowledge, automation environment, and testing standards. All too often, researchers focus on different technical aspects of software testing without taking into account the human aspects of different test activities. Humans involved in the creation of test cases are basically trying to solve certain problems that have been posed to them (e.g., assignments from test managers) or recognized on their own (e.g., the need for additional confidence in releasing a certain feature). Despite the diversity of these testing problems and goals, the ways people go about solving them show a number of common characteristics to general problem solving \cite{medin1992cognitive}. Problem solving is a very important topic of research in cognitive science but also in applied fields such as artificial intelligence. Over the years many researchers \cite{davidson2003psychology} have delved into this topic and have discovered a great deal of knowledge about how humans solve problems. We recognize here the need for examining and classifying the underlying characteristics of problem solving models when applied to software testing. 

\section{Related Work}
Behavioral Software Engineering (BSE) is defined as the study of cognitive, behavioral, and social aspects of software engineering as performed by individuals, groups, or organizations \cite{lenberg2015behavioral}. During the last couple of decades, many researchers have studied different cognitive aspects of quality assurance. For example, Hale et al.~\cite{hale1991towards} proposed a model of a programmer's cognitive skills during software maintenance. This cognitive process model for debugging is based on structural learning theory, which integrates both declarative models, such as a program comprehension model and problem-solving models. In a later study, Hale et al. \cite{hale1999evaluation} evaluated the proposed model through the use of verbal protocol analysis, which is accomplished through a controlled experiment, where the subjects are required to debug a program that contained an unknown error.

Robillard et al. \cite{robillard1998measuring} (in a multidisciplinary team composed of software engineers and cognitive psychologists) also studied the cognitive activities of software engineers. 
This study aimed to derive good practices based on observations and analyses of the processes that are typically used by software engineers. The experiment was captured on video, it was then transcribed, coded, and then further defined in categories, which helped in defining the cognitive behaviors. The results of this study shows that software review is composed of three types of
cognitive activities: review, synchronization, and elaboration of alternative solutions. 

Letovsky \cite{letovsky1987cognitive} studied the cognitive processes underlying program understanding, by focusing on events that occur in the order of seconds and minutes, such as identifying the intent behind a line of code. During analysis, a taxonomy for questions and conjectures was developed as well as a theory of the mental representations and processes that produced them. The questions were explained in terms of processes that evaluate the consistency and completeness of the human's developing mental model and the conjectures as a planning process operating on a variety of types of knowledge. 

Also related to program comprehension, Duraes et al.~\cite{duraes2016wap} studied the functional patterns activity in mapped brain regions associated with finding bugs. This study has been performed by using functional magnetic resonance imaging (fMRI). Their results confirmed that brain areas associated with language processing and mathematics are highly active during coding inspection and that there are specific brain activity patterns that can be related to decision-making during bug detection or suspicion. It seems that the right anterior insula may serve as a quality indicator of programmers' capacity to identify bugs when faced with a challenging piece of code.

Overall, there is a lack of studies concerning the behavioral, cognitive, and social aspects of software testing. Hale's model \cite{hale1991towards} seems to be the closest cognitive model related to software testing, but it is focusing on debugging tasks. There is a need to handle the issues involved in studying the behavioral aspects of human-based software testing. The aim of this paper is to touch on some of the cognitive aspects of software testing by investigating testing practices as problem solving.


 \begin{figure*}[ht!]
  	\centering   
    \includegraphics[width=0.65\textwidth]{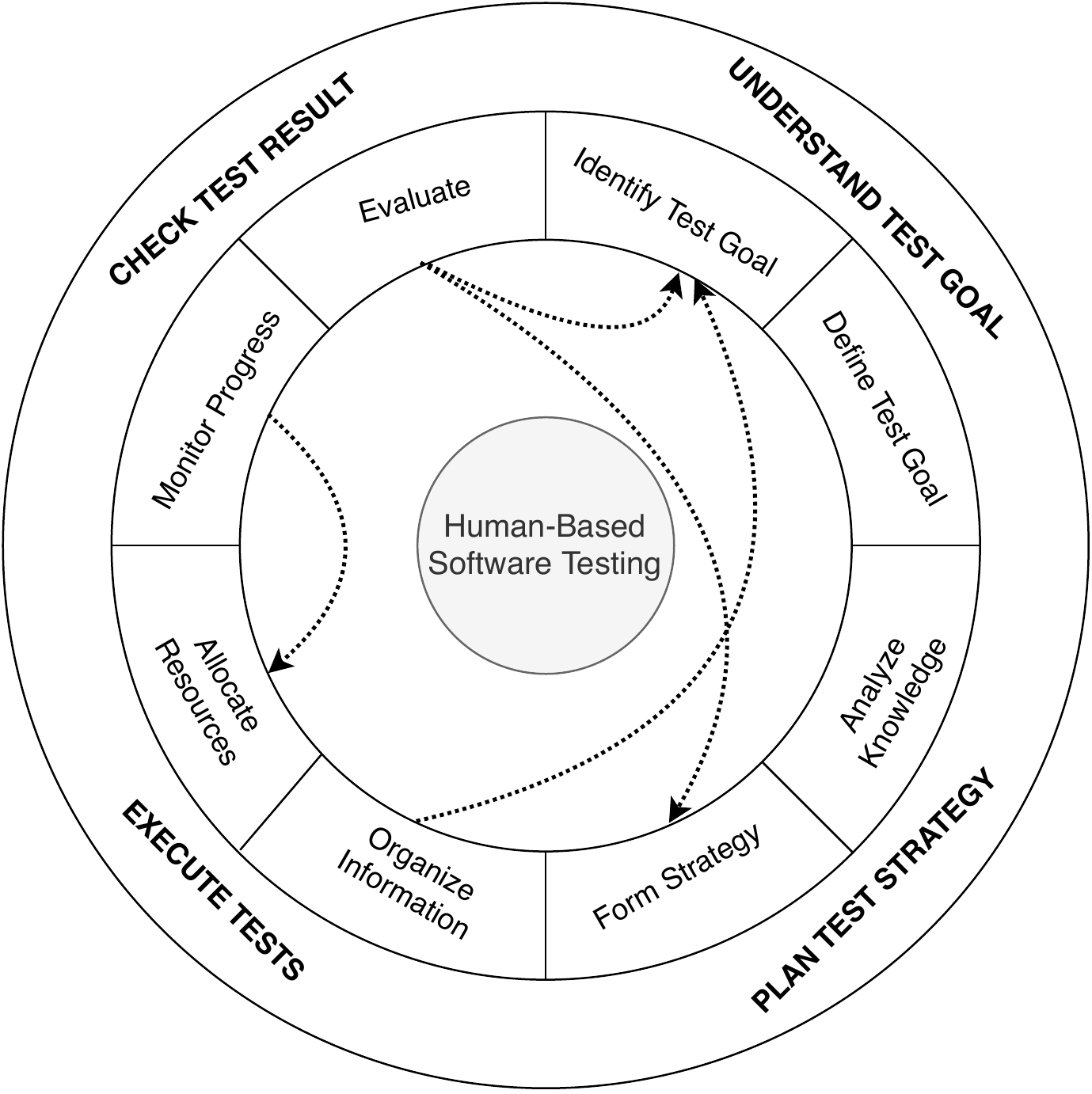}
 	\caption{The software testing process viewed as a cyclical problem solving model.}
 	\label{fig:model}
\end{figure*}

\section{Foundations for a cognitive model of software testing}
Many models representing the problem solving processes are devised principally from Polya's phases in solving mathematical problems~\cite{polya1957solve}. Several psychologists have also described the problem solving process as a cycle \cite{bransford1984ideal,hayes1989cognitive,pretz2003recognizing}. Problem solving occurs throughout life. A young child may be trying to figure out what a word means or how to solve a mathematical problem. An architect may be designing an apartment building, or a software tester is attempting to find certain effective test inputs using boundary value analysis. Despite the diversity of problems, examining these as a problem solving process, their commonalities and what they tell us about human thought and intelligence is important. Depending on the testing problem and how we mentally represent a test goal, different cognitive processes are used, some conscious and some unconscious. 

Traditionally, problem solving has been viewed in psychology and artificial intelligence research as determined by representation and search processes \cite{medin1992cognitive,newel1972human}. In this way, a problem is represented as a problem space, consisting of states and operators. A human or a program solves a problem when it finds a path from the initial goal to the goal state. In search-based software testing \cite{feldt2002biomimetic,mcminn2004search,afzal2009systematic,mcminn2011search}, much of the work has focused on understanding different search strategies that might be used for guiding the search towards specific test goals. 

More recently, problem representation has become an area of high interest in problem solving research \cite{pretz2003recognizing}. The representation seems to be at least as crucial in determining whether a problem is solved as to how the search is performed. Given these foundations on problem solving, we consider in the next section a cognitive model of software testing performed by human testers based on problem solving activities. 

Overall, this is quite a traditional view of problem solving consisting of fairly well-defined classical steps performed cyclically in a sequence. Other modern cognitive theories and models are more complex \cite{mustafic2019complex} since they handle different issues involved in studying the psychology of human problem solving such as developing expertise and skills \cite{feltovich2006studies}, insight, creativity and the neuroscience of problem solving \cite{davidson2003psychology}. To get a handle of all these issues in the software testing context, we first need a way of defining our terms and classifying testing problems. In the next section, we outline a software testing process devised from a classical problem solving model as a starting point for framing such investigations. Future enhancements can, possibly, benefit from considering more recent and holistic models proposed by later cognitive theorists and experimentalists.

\section{A cognitive model of software testing}
\label{acognitivemodel}
To get a handle on the issues involved in studying the (cognitive) psychology of software testing performed by humans, we need a way of defining and classifying the process of human-based software testing. In response to the need for a better understanding of the test creation process, this work explores the cognitive underpinnings used by testers in the test design phase. The model shown in Figure \ref{fig:model} is centered around our observation that test design and execution can be viewed as a problem solving process. In order to precisely describe this process, we need to better understand the cognitive processes associated with test design. While viewing it as a problem solving process is not the only psychological lens we could use, we argue it is a natural and fruitful one.

There are many models of problem solving (e.g., Schoenfeld's model \cite{schoenfeld2016mathematical}, Nunokawa's model \cite{nunokawa2005mathematical}, Jonassen's model\cite{jonassen1997instructional}, Hayes' model \cite{hayes1989cognitive}) developed on the basis of the four problem solving phases devised by Polya \cite{polya1957solve} for mathematical problem solving. In order to solve testing problems one would need to use a number of cognitive processes that are not directly covered by Polya's phases. In short, knowledge, skills, creativity and the cognitive resources needed to exploit them are involved in software testing. As a starting point, the software testing cycle viewed as a classical problem solving process consists of the following stages (mirrored in Figure \ref{fig:model}) in which the human tester must:
\begin{itemize}

\item \textit{Identify the Test Goal:} Recognize and identify the test goal as a problem that needs to be solved. According to Getzels \cite{getzels1982problem}, there are three main kinds of problems one can identify: those that are presented, those that are discovered, and those that are created. A presented test goal is one that is given to the tester directly and it is stated clearly (i.e., predefined criteria-based test goal). A discovered and\slash or created test goal, however, is one that first has to be recognized. In such cases, testers might be using exploratory testing strategies and\slash or seek out new test goals by using their intuition and experience.

\item \textit{Define the Test Goal:} Define and understand the test goal mentally and what the associated test cases must and should do. The test goal definition is the aspect of testing in which the scope and the test goals are clearly stated and described. A test goal provides the tester with a set of givens. Faced with these givens, a tester applies some kind of operations to reach the goal state (e.g., creation of a test case fulfilling the test goal). A test goal may be represented in a variety of ways, for example, visually or verbally. For instance, in achieving pairwise coverage \cite{ammann2016introduction}, one would need to define and represent the goal as the task to create all possible pairs of parameter values that can be covered by at least one test case.

\item \textit{Analyze Knowledge:} Organize his or her testing knowledge about the test goal. Everyone approaches a problem situation with a unique knowledge base. For somebody with knowledge in test design techniques, analyzing its prior knowledge involves using different kinds of operators depending on the test goal type and how we mentally represent it (e.g., mathematical operators). In order to create test cases, we would need to use general skills such as inferencing, case-based reasoning, abstraction and generalization for organizing the information from different steps. At an even more general level, there are metacognitive skills related to motivation and allocating cognitive resources such as attention and effort that one needs to use. In addition, one would need to use domain knowledge, which may include: electrical, mathematical and computer science concepts as well as programming concepts, rules and principles. 

\item \textit{Form Strategy:} Develop a solution strategy for creating the necessary test cases using certain operators. These operators are mental representations of the actions a tester can perform on the givens. For example certain calculations need to be done using mental operators. The set of operations you perform to get to the goal state constitutes the set of operations needed to create test cases fulfilling a certain test goal. In many cases, the operators are not specified in the test goal, but we can infer them from our prior knowledge (e.g., mathematical operators like division and multiplication, mental operators like calculations). 

\item \textit{Organize Information and Allocate Resources:} Organize information, allocate mental and physical resources for creating and executing test cases. For example, testers can use test automation for embedding the test values in executable scripts and allocate certain computing resources for executing test cases. If tests are run by hand, testers will allocate physical resources and record the results.

\item \textit{Monitor Progress:} Monitor his or her progress toward the test goal. This step monitors the results of test creation and execution. For test execution one will use test oracles embedded into scripts or manually monitor when the correct output cannot be easily encoded. 

\item \textit{Evaluate.} Evaluate the test cases for accuracy. When finding that the test goal is not met, we analyze the test goal and then make corrections. We follow the set of operations to see if the test cases really do not check the test goal.
\end{itemize}

The testers begin the testing process by analyzing the test goal, breaking it into manageable pieces, and developing a general solution for each piece called a test case. The solutions to the pieces are collected together to form a test suite that solves the original problem (i.e., the identified test goal). The testing cycle is descriptive and does not imply that all test case creation proceeds sequentially through all these steps in this order. In practice, experienced testers are those who are flexible. In many cases, once the cycle is completed, the steps are usually giving rise to a new test goal, and then the steps need to be repeated.

\section{What is a Testing Goal?}

Before discussing how one would investigate how people design test cases, we define what is meant by a test problem. For our purposes, a tester has a problem to solve when she wants to attain some test goal. We consider a test problem to have four aspects: test goals, assumptions, means to attain the goal and obstacles. The test goal is some state for which some criteria can be applied to assess whether the test problem has been solved. To give an example, the goal might be to check if a certain requirement has been implemented correctly or to find a way to crash the user interface.

Even if different test goals have common aspects, figuring out how to find test inputs to cover all branches in a program seems very different from figuring out how to find interface bugs in the same program. Clearly, these two test goals have many differences. A crucial activity in understanding problem solving is to analyze which test goal differences are important and which are not.

There are two classes of test goals that map directly on the generic problems solved by humans:
\begin{itemize}
\item \textit{Well defined test goals} (e.g., coverage criteria, boundary-value analysis, category partitioning) have completely specified initial conditions, goals and means of attaining the goal. Many coverage criteria are well defined. For example, creating tests for covering program branches is usually well defined. The goal is some well specified coverage score. Thus, you will know when you have attained your goal. Finally, the means of attaining the goal are by exercising the different branches in the program.
\item \textit{Ill defined test goals} (e.g., stress test goals, fault-based testing) have some aspects that are not completely specified. The problem of finding faults is clearly ill defined. Even if you know how to tell whether a fault is discovered, you wouldn't know exactly what to do to try to achieve this goal in every specific situation.
\end{itemize}

Test problems differ on how well defined they are and we can consider that this categorization is a continuum of problems rather than a dichotomy.

\section{Investigating Software Testing Practices as Problem Solving}
\label{methodsinvestigating}
In order to build knowledge on how testers are solving test problems by creating test cases, it is useful to consider the methods used in problem-solving research in cognitive science. Three methods are often used in problem-solving research \cite{pretz2003recognizing}: intermediate products, verbal protocols, and software simulations. 

\subsection{Intermediate Products}
Getting intermediate products \cite{medin1992cognitive} means that instead of recording and analyzing only the created test cases, we observe some of the work the subject does in getting these test cases. If we are interested in how people create test cases for finding logical bugs, we collect information about the various steps they make in getting to the goal. If we are interested in testing for requirement coverage, we collect and analyze the models, equations and other information the subject writes down in the course of problem solving. The resulting intermediate products provide finer constraints and possible explanations. 

\subsection{Verbal Protocols} 
The second method often used in problem solving research is a verbal protocol \cite{ericsson1998protocol}. The most common way to collect such data is to ask the subjects to think aloud as they go about solving the problem. The idea behind this measure is to provide information about the course of the problem solving. Verbal protocols can be used in software testing research as direct evidence for some hypothesis or to generate new ideas that can be tested by other methods. 

\subsection{Simulations and Search-Based Testing}
A common goal of problem solving research is to build a simulation that is meant to mimic the problem-solving process as revealed by the intermediate steps \cite{medin1992cognitive}. In testing, automatic search-based test generation \cite{mcminn2011search} can be used to mimic how testers solve testing problems, to the extent that problem solving can be regarded as information processing.

\section{Experiment Design} 
To test our hypothesized problem solving model of test design we propose an experiment in which we obtain data using one of the methods outlined in Section \ref{methodsinvestigating} (i.e., verbal protocols \cite{medin1992cognitive}). The goal of the experiment is to verify the test design cognitive model outlined in Section \ref{acognitivemodel} by determining whether the stages of this process cycle are shown during a test design session. In addition, we will identify the patterns of the cognitive processes that are used by classifying the series of the cognitive actions that the subjects will undertake during the problem-solving process. It is also important to identify the comprehension strategies used in the experiment by verifying whether specific comprehension strategies, such as the usage of prior knowledge or questioning is used during the experiment.

  \subsection{Experiment Material} 
 The experiment materials used for the conduct of the experiment are the informed consent, the survey used for the selection of the participants, the software programs used for test design as well as the software specification documents. 
  
 The informed consent gives the participants the basic idea of the experiment and what their participation involves. The form contains the purpose of this research. In this study, we use the information that the participants provide to map the steps and sequence by which testers perform test activities and evaluate whether these steps correspond to existing models. In addition, to be recruited for this study, the participants should be comfortable talking about their daily routine in creating test cases, technology usage, and thoughts about all aspects influencing their test creation process. We need them to provide detailed information about their thought process during testing. Their anonymity needs to be strictly maintained. The data that will be collected will be labeled with an anonymous participant ID. Participants will remain anonymous, but researchers will refer to participants (if at all) by an ID in all subsequent analysis artifacts. Any identifying piece of information will be erased. The risks of participation are intended to be none or minimal. However, because they will provide detailed information about their work, there are concerns about privacy. To mitigate this risk, the subjects can choose what information they are comfortable revealing. No one except the researchers will be allowed to see any of the information they provide. All electronic data and data collected as part of the study will be kept on an external hard drive and stored in a locked cabinet. It is expected to issue papers and presentations describing this research. Public presentations of the results will essentially present the results in an aggregate form. In cases where the individual participant data is disclosed, such as comments or quotes, we will ensure that the selected data does not suggest participant identities.
 
 \begin{table*} [htp]
    \centering
    \begin{tabular}{|l|l|l|}
    \hline
    \textbf{Predicate} & \textbf{Argument} & \textbf{Notes}  \\
    \hline  
    \hline
    Identify Test Goal & \textless knowledge\textgreater & Explicit knowledge \\ 
    \hline
    Define Test Goal  & \textless knowledge\textgreater & Explicit knowledge\\
    \cline{3-3} & & Implicit knowledge \\ 
    \hline
    Analyze Knowledge & \textless knowledge\textgreater &  Explicit knowledge\\
    \cline{3-3} & & Implicit knowledge\\ 
    \hline
    Form Strategy &\textless strategy\textgreater & Document and code inspection/search\\
    \cline{3-3} & & Examination of the program output \\
    \cline{3-3} & & Simulation of program execution\\
    \hline
    Organize Information & \textless knowledge\textgreater & Explicit knowledge\\
    \cline{3-3} & & Implicit knowledge \\ 
    \hline
    Allocate Resources & \textless action\textgreater &  Writing of test data\\
    \cline{3-3} & & Writing of expected outputs \\ 
    \hline
    Monitor Progress & \textless action\textgreater & Execution of test cases \\ \hline
    Evaluate & \textless what\textgreater & Verification of outputs\\
    \cline{3-3}& & Program Representation \\ \hline
    Silence & \textless manner\textgreater & Silence / prompts\\
    \cline{3-3} & & No coding \\\hline
    Positive / Negative feelings & \textless manner\textgreater & Reactions\\ [1ex] 
    \hline
    \end{tabular}
    \caption{Test Design Encoding Scheme.}
    \label{encondingscheme}
\end{table*}
 
  \subsection{Tasks of the participants}  

\begin{enumerate}
    \item Participants will read the specification document containing the necessary information of the program under test. 
    \item Participants will test the program under test to the best of their abilities. They will need to create test cases and systematically test the program to increase their confidence that the software works and to find bugs. We are specifically interested in their thinking process and experiences in creating new test cases. It is essential for us to understand what the participants are thinking about as they work on the test creation task, starting with identification and understanding the test goal or purpose until they execute the created test case.
        \item Participants will complete both tasks 1 and 2 based on the Verbal Protocol Analysis (VPA) method.
    \begin{itemize}
        \item While executing task 1, subjects will articulate all their thoughts and opinions that they might have while reading the document.
        \item While executing task 2, subjects will express everything that they are doing, e.g., why they are creating specific test cases or when they will stop testing a particular program functionality.
        \item It is required from the participants to talk about everything that they are thinking from the moment they start the task until they have completed it; they must continuously talk and clearly state the order of their thinking and events.
    \end{itemize}
\end{enumerate}

  \subsection{Initial Procedure and Recording} 
    \label{procedure}
 The participants will enter the room one by one according to a certain schedule, where the only individuals in the area of the experiment will be the subject and one researcher. The subject will be provided with a computer where she will create the test cases and a specification document containing the program's information. A video camera will be installed near the subject where the camera's frame will capture the participant's front side and partly the researcher. The camera will also capture the audio of the experiment. Nevertheless, an extra audio-capturing device will be used in case any noise interference is present.
 
     \subsection{Analysis Procedure} 
 Data analysis will be based on the verbal protocol encoding approach proposed by Ericsson et al. \cite{ericsson1998protocol}. The behavior of the participants will be captured on video and audio format, and is followed by a transcription of their recordings. This data will then be divided into segments (e.g., phrases, sentences, clauses). These segments will then be encoded. The encoded data will be analyzed via two different steps, the content and pattern analysis. Content analysis provides an independent assessment of whether the process steps articulated in the test design cognitive model represent the problem-solving activities exhibited by the participants. The pattern analysis examines the existence of patterns and relationships via a within-subject analysis and between-subject analysis based on the information gathered by the content analysis. The within-subject identifies for each subject process the series of steps that may occur more often than others, while the between-subject analysis identifies the process step series that often occur between the participants.
 
 \begin{table*}[htp]
    \centering
    \begin{tabular}{|l|p{4cm}|c|}
    \hline
    \textbf{Predicate} & \textbf{Participants Demonstrating Step} & \textbf{Average Mean \% of Total Activities}\\
    \hline
    \hline
     Identify Test Goal & \centering5/5 & 7.1 \% \\
     \hline
     Define Test Goal & \centering 5/5 & 5.4\% \\
     \hline
     Analyze Knowledge & \centering 5/5 & 13.1 \% \\
     \hline
     Form Strategy & \centering 5/5 & 16.1 \% \\
     \hline
     Organizing Information & \centering 5/5 & 10.3 \% \\
     \hline
     Allocate Resources & \centering 5/5 & 5.1 \% \\
     \hline
     Monitor Progress & \centering 5/5 & 2.6 \% \\
     \hline
     Evaluate & \centering 5/5 & 6.3 \% \\
     \hline
     Silence & \centering 5/5 & 27.6 \% \\
     \hline
     Positive and Negative Feelings & \centering 4/5 & 1 \% \\
     \hline
     N/A & \centering 5/5 & 5.3 \% \\
     \hline
    \end{tabular}
    \caption{Analysis of the Observed Process Steps}
    \label{analysistotal}
\end{table*}
 
 The adopted coding scheme (shown in Table \ref{encondingscheme}) is based on a predicate calculus notation in which predicates represent the hypothesized test design process steps. Analysis will be executed based on an encoding scheme shown in Table \ref{encondingscheme}. Each of the segments will refer to a specific predicate, and after all the segments have been encoded, the analysis can start. Table \ref{encondingscheme} states each of the predicates used in the experiment. 
\noindent Itkonen et al. \cite{itkonen2012role} have reported the results of an experiment identifying different types of knowledge that a tester may use when designing a test case. Such knowledge can be categorized into two types: explicit knowledge and implicit knowledge.~Explicit knowledge concerns with the systems knowledge or the documentation (i.e., information used regarding the software or documentation), while implicit knowledge is focusing on the domain knowledge and the experience that a tester has (e.g., practical or conceptual knowledge of the subject matter and tools). Given that these types of knowledge are being used, various predicates in the Encoding Scheme have a \textless knowledge\textgreater~argument. \textit{Identify Test Goal} is using the explicit knowledge of the testers in regards to the documentation or the code. \textit{Define Test Goal} is using both the explicit and implicit knowledge of the testers. While defining the test goal, the tester needs to understand the code entirely based on prior experience as well. \textit{Analyze knowledge}, as the name suggests, assumes that a tester needs both types of knowledge to analyze the necessary information to form a strategy for the creation of test cases. Another predicate that uses both types of knowledge is the \textit{Organizing Information}, as the tester needs to have all the possible information during reasoning. \textit{Form Strategy} contains the argument \textless strategy\textgreater, which refers to searching through the code and test scripts in order to find information or to support current findings. Another way to create a strategy is through examining the output of the program, to understand if a certain test case is fulfilling the test goal. In some other cases, test cases are re-executed if the strategy needs to be fixed in some way. The \textless action\textgreater~argument mentioned in two different predicates in the scheme refers to two different actions: the predicates \textit{Allocate Resources} and \textit{Monitor Progress}. \textit{Evaluate} contains the \textless what\textgreater~argument, which deals with verifying the output of the program against a certain test case and checking if the test goal has been fulfilled. The other two predicates we use are \textit{Silence} and \textit{Positive / Negative Feelings}, both using the \textless manner\textgreater~argument. The \textit{Silence} refers to periods of time when no words are being spoken, while \textit{Positive / Negative Feelings} refers to the positive or negative reactions a tester might express during the entire test design procedure.

\section{Pilot Study Results}
As the study setting available to use was limited to a non-industrial environment and a physical space at M\"{a}lardalen University in Sweden, we restricted the pilot experiment to an academic environment. The behaviour of each subject was captured on audio and video tape. These recordings were transcribed by dividing the verbalizations into segments, each representing an idea or an action and then the segments were encoded. 
\subsection{Research Subjects and System-Under-Test}
Five students, all trained in software
testing at master level at Mälardalen University, participated in the pilot study. Based on the results of the survey, these participants had experience in programming using Java and test creation using JUnit. The subjects did not earned credits for participation or performed this experiment as part of any course work.

The program used in this experiment is a bowling game calculator. The code is written in Java, is 129 lines of code long, and it calculates the score of the rounds of a normal bowling game. The program has only one class which is composed of several methods. 

\subsection{Content Analysis}
In this pilot study, we focused on performing content analysis and providing initial evidence of each of the hypothesised test design process steps included in the a priori cognitive model representation.

The content analysis, summarized in Table \ref{analysistotal}, suggests that each step derived from the
cyclical problem solving model is consistently performed by each of the five human subjects when testing the bowling game program. Based on these initial confirmatory results, we can assume that the individual testing process steps adequately describe the actual testing activities exhibited by these subjects.

For an average of 12\% of the total testing time, the subjects identified and defined the test goal by understanding this problem and what a test case is supposed to perform. For 39\% of their time on average, the subjects attempted to analyze their knowledge regarding the test goal and planned different approaches on how to create test cases. Participants organized information via inferencing and case-based reasoning. When the necessary information had been gathered, the subjects started to allocate resources to create test scripts (5.1\% on average of the total testing time). For 9\% of their time on average, participants monitored and evaluated their progress. Out of all the five participants, four of them showed positive or negative feelings and reactions about the process of creating test cases. The N/A predicate in Table \ref{analysistotal} refers to behaviours that could not be coded. These were attempts to interact with the researcher during the
testing task or verbalizing a physical
activity which is not related to the actual testing process.

In this pilot study, we have not performed any systematic analysis in which to check if the hypothesized sequence by which the testing process steps are followed by all participants. Some of these sequences are only partially supported by some of the data collected in this experiment. More studies are needed to confirm the specific sequence of steps used during the testing process.

\section{Threats to Validity}
We acknowledge here that we are employing content analysis of verbal protocols that are used to test the problem solving model of software testing. For this purpose, the encoding scheme is formally defined before the experiment and this context greatly constrains the range of possible interpretations of the content. Since the data is gathered to test a model, there is a risk of over-fitting the steps actually taken by participants. To counter this, the data interpretation and encoding scheme is based on a generic and simple process.

All of our subjects are students and have limited professional software development experience. This fact has been shown to be of somehow minor importance in certain conditions in a study by H\"ost et al. \cite{host2000using} with software engineering students being good substitutes in experiments for software professionals. 

Our study has focused on one rather small Java program. Nevertheless, we argue that this program is representative of software which testers might encounter in their career. 

\section{Conclusions}
We have shown that models of 
problem solving from cognitive psychology can help in understanding the cognitive processes of software testers. The software testing model developed here is cyclical and can accommodate both simple and complex test goals of different levels of clarity and sophistication.

For the research community, the model proposed here provides a reference for future investigations. In general, more and more specific questions concerning testers' cognitive processes in software testing can be posed. Specifically, the process by which a tester identifies and defines a test goal is not fully understood, nor is the process by which testers form strategies for creating test cases. The selection of search strategies could vary with testing experience, but little is known about this. Our model could help designing relevant experiments and organize observations in case studies. Likewise, how a tester switches between selecting test goals, designing test cases and inputs, and executes test cases has received little attention. 

The model also provides a framework from which to investigate tester knowledge and expertise and it is the first step in evaluating and refining a testers' cognitive process model for software testing. 

The results of the pilot evaluation support the postulated cognitive model in an academic setting in an Eclipse environment with a Java program. Our results indicate that the steps of this model were observed when subjects tested an already developed program. More experimentation is needed to investigate real testing
situations in which testers use more complex test goals and strategies. In addition, there is a need to explore how they select and switch test goals, create test cases and use different comprehension
strategies. 

\balance

\section*{Acknowledgment}
This work is partially supported by the Software Center Project 30 on Aspects of Automated Testing and by the Swedish Innovation Agency (Vinnova) through the XIVT project. The authors would also like to thank Per Erik Strandberg for his valuable comments on this work.
\bibliographystyle{IEEEtran}
\bibliography{acmart} 

\end{document}